\documentclass[journal]{IEEEtran}

\usepackage{cite}
\usepackage{algorithmic}
\usepackage{graphicx}
\usepackage{algorithm,algorithmic}
\usepackage{hyperref}
\hypersetup{hidelinks=true}
\usepackage{textcomp}

\usepackage{amsmath,amssymb,amsfonts,mathtools}
\usepackage{textcomp}
\usepackage{xcolor}
\usepackage{siunitx}
\usepackage{makecell}
\usepackage{array}
\usepackage{booktabs}
\usepackage{soul}
\usepackage[nolist]{acronym}
\usepackage{svg}
\usepackage{mathdots}
\usepackage{pgfplots}
\usepackage{caption}
\usepackage{subcaption}
\usepackage{tabularx}
\usepackage{multirow}
\usepackage{arydshln}  
\usepackage{booktabs}
\usepackage{comment}
\usepackage{bm}

\usepackage{mdframed}

\newcolumntype{C}{>{\centering\arraybackslash}X}

\usepackage{tikz}
\usepackage{tikz-3dplot}
\usetikzlibrary{positioning,decorations.pathreplacing,calligraphy}

\renewcommand{\vec}{\mathbf}

\newcommand{\todo}[1]{\textcolor{red}{#1}}

\mathchardef\mhyphen="2D

\usepackage{orcidlink}

\newcommand{\orcidauthorB}{\orcidlink{0000-0003-1460-1061}} 
\newcommand{\orcidauthorC}{\orcidlink{0000-0002-5304-4213}} 
\newcommand{\orcidauthorH}{\orcidlink{0000-0001-5877-1439}} 
\newcommand{\orcidauthorL}{\orcidlink{0000-0002-4806-9838}} 
\newcommand{\orcidauthorO}{\orcidlink{0009-0003-2171-8046}} 
\newcommand{\orcidauthorP}{\orcidlink{0000-0002-0318-6133}} 
\newcommand{\orcidauthorQ}{\orcidlink{0000-0002-9413-9083}} 

\definecolor{Color1}{HTML}{00ff29}
\definecolor{Color2}{HTML}{00ffd6}
\definecolor{Color3}{HTML}{00ff7e}
\definecolor{Color4}{HTML}{38ff00}
\definecolor{Color5}{HTML}{00d4ff}
\definecolor{Color6}{HTML}{55ffaa}

\newcommand{\markerbox}[1]{%
  \fcolorbox{black}{#1}{\parbox{2em}{\mathstrut}}%
}

\newcommand{\mybar}[1]{%
    {\color{black}\rule{0.6pt * #1}{6pt}}%
    
}

\begin{document}

\title{MASSLOC: A Massive Sound Source Localization System based on Direction-of-Arrival Estimation and Complementary Zadoff-Chu Sequences}

\author{Georg~K.J.~Fischer\orcidauthorB, Thomas~Schaechtle\orcidauthorC, Moritz~Schabinger\orcidauthorO, Alexander~Richter\orcidauthorQ, Ivo~Häring\orcidauthorP, Fabian~Höflinger\orcidauthorH, and Stefan~J.~Rupitsch\orcidauthorL, \IEEEmembership{Member, IEEE}
\thanks{This work was partially supported German Federal Ministry of Economic Affairs and Climate Action (BMWK) under grant FKZ 03EE3066D Verbundvorhaben LoCA and by the German Federal Ministry of Education and Research (BMBF) under grant FKZ 13N16818 Verbundprojekt FreiburgRESIST.}
\thanks{Georg~K.J.~Fischer, Thomas Schaechtle, Alexander Richter, Ivo Häring and Fabian~Höflinger are with the Fraunhofer Institute for Highspeed Dynamics, Ernst-Mach-Institute (EMI), Freiburg, Germany, E-Mail: georg.fischer@emi.fraunhofer.de.}
\thanks{Thomas~Schaechtle, Moritz Schabinger, Alexander Richter, Fabian~Höflinger and Stefan~J.~Rupitsch are with the Department of Microsystems Engineering (IMTEK), University of Freiburg, Germany.}}

\markboth{Journal of \LaTeX\ Class Files,~Vol.~14, No.~8, August~2015}%
{Shell \MakeLowercase{\textit{et al.}}: Bare Demo of IEEEtran.cls for IEEE Journals}

\maketitle

\begin{abstract}
Acoustic indoor localization offers the potential for highly accurate position estimation while generally exhibiting low hardware requirements compared to Radio Frequency (RF)-based solutions. Furthermore, angular-based localization significantly reduces installation effort by minimizing the number of required fixed anchor nodes. In this contribution, we propose the so-called MASSLOC system, which leverages sparse two-dimensional array geometries to localize and identify a large number of concurrently active sources. Additionally, the use of complementary Zadoff–Chu sequences is introduced to enable efficient, beamforming-based source identification. These sequences provide a trade-off between favorable correlation properties and accurate, unsynchronized direction-of-arrival estimation by exhibiting a spectrally balanced waveform. The system is evaluated in both a controlled anechoic chamber and a highly reverberant lobby environment with a reverberation time of 1.6\,s. In a laboratory setting, successful direction-of-arrival estimation and identification of up to 14 simultaneously emitting sources are demonstrated. Adopting a Perspective-n-Point (PnP) calibration approach, the system achieves a median three-dimensional localization error of 55.7\,mm and a median angular error of 0.84\,° with dynamic source movement of up to 1.9\,ms\textsuperscript{-1} in the challenging reverberant environment. The multi-source capability is also demonstrated and evaluated in that environment with a total of three tags. These results indicate the scalability and robustness of the MASSLOC system, even under challenging acoustic conditions.

\end{abstract}


%
\IEEEpeerreviewmaketitle

\begin{acronym}[JSONP]\itemsep0pt
    \acro{AN}{Anchor Node}
    \acro{ANN}{Artificial Neural Network}
    \acro{AGV}{Autonomous Guided Vehicle}
    \acro{AP}{Access Point}
    \acro{AoA}{Angle-of-Arrival}
    \acro{AoD}{Angle-of-Departure}
    \acro{BLE}{Bluetooth Low-Energy}
    \acro{BT}{Bluetooth}
    \acro{CCF}{Cross-Correlation Function}
    \acro{CDF}{Cumulative Distribution Function}
    \acro{CFO}{Carrier Frequency Offset}
    \acro{CIR}{Channel Impulse Response}
    \acro{CNN}{Convolutional Neural Network}
    \acro{COTS}{Commercial Off-the-shelf}
    \acro{CRLB}{Cramér-Rao Lower Bound}
    \acro{CSI}{Channel State Information}
    \acro{CTE}{Constant Tone Extension}
    \acro{CW}{Continuous Wave}
    \acro{DA}{Directive Antenna}
    \acro{DL}{Downlink}
    \acro{DL-AoD}{Downlink AoD}
    \acro{DULA}{Double-ULA}
    \acro{DoA}{Direction-of-Arrival}
    \acro{DoD}{Direction-of-Departure}
    \acro{DoP}{Dilution of Precision}
    \acro{ESPRIT}{Estimation of Signal Parameters via Rotational Invariance Techniques} 
    \acro{EToA}{Elapsed Time between two Times-of-Arrivals}
    \acro{FCC}{Federal Communications Commission}
    \acro{FHSS}{Frequency-Hop Spread Spectrum}
    \acro{FIM}{Fisher Information Matrix}
    \acro{FP}{Fingerprint}
    \acro{FR1}{Frequency Range 1}
    \acro{FR2}{Frequency Range 2}
    \acro{FSA}{Frequency-Scanned Antenna}
    \acro{FTM}{Fine Timing Measurement} 
    \acro{FoV}{Field-of-View} 
    \acro{GCC-PHAT}{Generalized Cross-Correlation Phase Transform} 
    \acro{GDoP}{Geometrical Dilution of Precision}
    \acro{GFSK}{Gaussian Frequency Shift Keying} 
    \acro{GMM}{Gaussian Micture Model}
    \acro{GNSS}{Global Navigation Satellite System}
    \acro{GPS}{Global Positioning System}
    \acro{GRU}{Gated Recurrent Unit}
    \acro{IPS}{Indoor Positioning System}
    \acro{IoT}{Internet-of-Things}
    \acro{JADE}{Joint Angle and Delay Estimation}
    \acro{LBS}{Location-Based Service}
    \acro{LPP}{LTE Positioning Protocol}
    \acro{LS}{Least-Squares}
    \acro{LTE}{Long Term Evolution}
    \acro{LoS}{Line-of-Sight}
    \acro{M2M}{Machine-to-Machine}
    \acro{MAE}{Mean Absolute Error}
    \acro{MCU}{Microcontroller Unit}
    \acro{MP}{Monopulse}
    \acro{MPI}{Multipath Influence}
    \acro{MUSIC}{MUltiple SIgnal Classification} 
    \acro{NIC}{Network Interface Card}
    \acro{NLoS}{Non-Line-of-Sight}
    \acro{NURA}{Non-Uniform Rectangular Array}
    \acro{PA}{Phased Array}
    \acro{PDDA}{Propagator Direct Data Acquisition}
    \acro{PDoA}{Phase Difference of Arrival}
    \acro{RADAR}{Radio Detection and Ranging}
    \acro{RF}{Radio Frequency}
    \acro{RFID}{Radio Frequency Identification}
    \acro{RIS}{Reconfigurable Intelligent Surface}
    \acro{RMSE}{Root Mean Square Error} 
    \acro{RSS}{Received Signal Strength}
    \acro{RSSI}{Received Signal Strength Indicator}
    \acro{RTLS}{Real-Time Localization System}
    \acro{RTT}{Round Trip Time}
    \acro{SA-AoA}{Single Antenna AoA}
    \acro{SAA}{Switched Antenna Array}
    \acro{SAL}{Single Anchor Localization}
    \acro{SAR}{Synthetic Aperture Radar}
    \acro{SBA}{Switched Beam Antenna}
    \acro{SDR}{Software-Defined Radio}
    \acro{SNR}{Signal-to-Noise Ratio}
    \acro{TDoA}{Time-Difference-of-Arrival}
    \acro{TPSN}{Timing-sync Protocol for Sensor Networks}
    \acro{TWR}{Two Way Ranging}
    \acro{ToA}{Time-of-Arrival}
    \acro{ToE}{Time-of-Emission}
    \acro{ToF}{Time-of-Flight}
    \acro{UCA}{Uniform Circular Array}
    \acro{UE}{User Equipment}
    \acro{UL}{Uplink}
    \acro{UL-AoA}{Uplink AoA}
    \acro{UL-TDoA}{Uplink TDoA}
    \acro{ULA}{Uniform Linear Array}
    \acro{URA}{Uniform Rectangular Array}
    \acro{USRP}{Universal Software Radio Peripheral} 
    \acro{UWB}{Ultra-Wideband}
    \acro{VLC}{Visible Light Communication}
    \acro{VLP}{Visible Light Positioning}
    \acro{WLS}{Weighted Least Squares}
    \acro{Wi-Fi}{Wireless Fidelity}
    \acro{IMU}{Inertial Measurement Unit}
    \acro{MDS}{Multidimensional Scaling}
    \acro{EIRP}{Equivalent Isotropically Radiated Power}
    \acro{MSE}{Mean Squared Error}
    \acro{TSDN}{Transformer-based Signal Denoising Network}
    \acro{IC}{Integrated Circuit}
    \acro{WLAN}{Wireless Local Area Network}
    \acro{MIMO}{Multiple Input Multiple Output}
    \acro{DFL}{Device-free Localization}
    \acro{PDR}{Pedestrian Dead Reckoning}
    \acro{SPBF}{Source Separation and Beamforming}
    \acro{CRC}{Cyclic Redundancy Check}
    \acro{PDU}{Protocol Data Unit}
    \acro{mmWave}{Milimeter Wave}
    \acro{ERP}{Enterprise Resource Planning}
    \acro{NRPPa}{New Radio Positioning Protocol A}
    \acro{PD}{Photodiode}
    \acro{PSD}{Position Sensitive Device}
    \acro{QADA}{Quadrant Angular Diversity Aperture}
    \acro{UL-SRS}{Uplink Sounding Reference Signal}
    \acro{DL-PRS}{Downlink Positioning Reference Signal}
    \acro{gNB}{Next-Generation NodeB}
    \acro{BS}{Base Station}
    \acro{RSRP}{Reference Signal Received Power}
    \acro{3GPP}{3rd Generation Partnership Project}
    \acro{GSM}{Global System for Mobile Communications}
    \acro{ISM}{Industrial, Scientific and Medical}
    \acro{FMCW}{Frequency-Modulated Continuous-Wave}
    \acro{OMP}{Orthogonal Matching Pursuit}
    \acro{SRP}{Steered Response Power}
    \acro{MVDR}{Minimum Variance Distortionless Response}
    \acro{LASSO}{Least Absolute Shrinkage and Selection Operator}
    \acro{UHF}{Ultra High Frequency}
    \acro{FPI}{First Path Index}
    \acro{OFDM}{Orthogonal Frequency-Division Multiplexing}
    \acro{FNN}{Feedforward Neural Network}
    \acro{LR-WPAN}{Low-Rate Wireless Personal Area Network}
    \acro{PCB}{Printed Circuit Board}
    \acro{PRISMA}{Preferred Reporting Items for Systematic reviews and Meta-Analyses}
    \acro{FiRa}{Fine Ranging}
    \acro{IMU}{Inertial Measurement Unit}
    \acro{IQ}{In-Phase Quadrature}
    \acro{EPS}{Enterprise Management Systems}
    \acro{API}{Application Programming Interface}
    \acro{ZC}{Zadoff-Chu}
    \acro{FPGA}{Field Programmable Gate Array}
    \acro{HPS}{Hard Processor System}
    \acro{BO}{Band-Overlap}
    \acro{BS}{Band-Separated}
    \acro{C-ZC}{Complementary Zadoff-Chu}
    \acro{MoCap}{Motion Capture}
    \acro{P-n-P}{Perspective-n-Point}
    \acro{SBL}{Sparse Bayesian Learning}
    \acro{DS}[D\&S]{Delay-and-Sum}
\end{acronym}
\section{Introduction}
\begin{figure}
    \centering
    \includegraphics[width=1\linewidth]{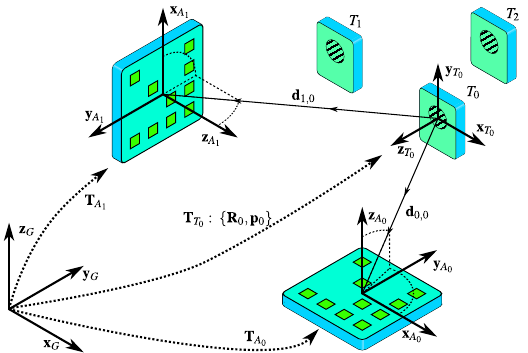}
    \vspace{-2mm}
    \caption{Illustration of the general system model for the angular-based indoor localization system. The figure shows two microphone arrays, \(A_0\) and \(A_1\), with their local coordinate frames (\(\vec{x}_{A_0}, \vec{y}_{A_0}, \vec{z}_{A_0}\) and \(\vec{x}_{A_1}, \vec{y}_{A_1}, \vec{z}_{A_1}\)) and multiple transmitting sources (\(T_1, T_2, T_3\)) emitting orthogonal signals. The distances \(d_{0,0}\) and \(d_{1,0}\) represent the paths from the sources to the arrays. The global coordinate frame (\(\vec{x}_G, \vec{y}_G, \vec{z}_G\)) and transformations (\(T_{A_0}\), \(T_{A_1}\), etc., with rotation matrix \(\vec{R}_i\) and translation vector \(\vec{p}_i\)) between the local and global coordinate systems are depicted.}
    \label{fig:system-model}
\end{figure}

To this day, there is no well-established or widely available indoor localization system comparable in ubiquity and compatibility to outdoor \acp{GNSS}. Nevertheless, indoor positioning services are already heavily used in industrial contexts, e.g., for tracking of assets, monitoring production processes, or guiding \acp{AGV} in logistics. Moreover, a growing number of consumer applications exists, such as navigation in shopping malls or museums. Despite this demand, no universal indoor localization system has been established that is supported across a broad range of devices and available as a general-purpose solution \cite{Zafari.2019,Girolami.2024,Fischer.2022}.

One of the main reasons for this is the high installation and infrastructure cost required by many systems, particularly those relying on time-delay measurements. Techniques such as \ac{ToF}, \ac{TDoA}, or \ac{ToA} often require a large number of fixed anchor nodes distributed throughout the environment. In addition, these systems impose strict synchronization requirements, sometimes even between mobile devices and anchors. The level of synchronization required is determined by the underlying measurement principle: for instance, acoustic systems set lower requirements in terms of synchronization than RF-based systems, due to the much slower propagation speed of sound in air \cite{Zafari.2019}.

Acoustic systems offer further benefits. They can achieve high localization accuracy using low-cost, off-the-shelf components and are suitable for deployment in \ac{RF} sensitive environments, such as laboratories or hospitals. Furthermore, they offer excellent device compatibility, as virtually all commercially available smartphones can be integrated through their built-in speakers and microphones.

Among acoustic techniques, angular-based systems provide additional information about the spatial relationship between sender and receiver, particularly the \ac*{DoA}. Systems capable of estimating full 3D direction vectors can theoretically perform full 3D localization with only two anchor nodes, whereas TDoA systems typically require at least four fixed anchors. Moreover, direction-based systems do not require tight time synchronization between receivers. Instead, it is sufficient that measurements are temporally associated, which imposes significantly more relaxed timing constraints \cite{Fischer2025}.

However, these benefits come with certain challenges. The hardware and software complexity of direction-based systems is considerably higher, as they must extract more information from the received signals. Although fewer anchor nodes are needed, the number of calibration parameters per node doubles. While time-delay systems only require the 3D position of the node, direction-based systems also require accurate knowledge of each node’s spatial orientation. These practical deployment aspects must be effectively addressed to fully leverage the advantages of direction-based localization, e.g. by the use of self-calibrating algorithms \cite{Gburrek2021,Shames2013,Bishop2009,Hahmann2022}.

In this contribution, we address several practical aspects of acoustic direction-based indoor localization. First, sparse microphone array geometries are evaluated, which can maintain high angular estimation accuracy while reducing the number of physical sensors and thereby enabling more compact devices and even supporting scenarios where more sources are resolved than there are sensors \cite{Pesavento2023,Aboumahmoud.2021}. Second, in contrast to most existing work, this system supports the low-latency tracking and localization of multiple concurrently active sources, even under signal collisions, enabled by a narrowband estimation approach \cite{Gabbrielli.2023b}.  To enable robust source identification, we introduce a novel set of complementary Zadoff-Chu waveforms. These sequences offer a trade-off between spectral symmetry for accurate, unsynchronized direction estimation and orthogonality, which is needed for distinguishing and identifying multiple concurrent emitters. \autoref{fig:system-model} provides an overview of the general system model and the symbolic notation used.

All these components are integrated into a unified localization framework based on beamforming and hypothesis testing. The proposed system is then evaluated through both controlled laboratory experiments and challenging, reverberant indoor environments with practical self-calibrating algorithms.

The primary contributions of this work are as follows:

\begin{itemize} 
    \item An acoustic angular-based \ac{IPS} is proposed, capable of handling multiple concurrently emitting sources using sparse arrays. 
    \item The proposal of a novel set of complementary Zadoff-Chu sequences, which are suited for the application to unsynchronized and concurrent direction estimation, as well as source identification.
    \item Comprehensive simulations are conducted across various parameter sets to guide the design of an optimal system. 
    \item The proposed system is evaluated experimentally in a laboratory setting and under real-world indoor conditions. 
\end{itemize}

The remainder of this work is structured as follows: \autoref{sec:rel-work} provides an overview of related work in the areas of acoustic and angular-based localization. \autoref{sec:loc-meth} introduces the system model and outlines the methodology for direction-of-arrival estimation using sparse arrays. \autoref{sec:sec-balancing} presents the design of orthogonal waveform sequences and their role in source identification. \autoref{sec:system-overview} briefly describes the system overview and hardware concept. \autoref{sec:exp} details the experimental setup and presents evaluation results under both controlled and real-world indoor conditions. Finally, \autoref{sec:conclusions} concludes this contribution and outlines potential directions for future research.
\section{Related Work}
\label{sec:rel-work}
The field of indoor localization and positioning is characterized by the use of a wide variety of technologies. To date, no ubiquitous and precise system, comparable to \ac{GNSS} for outdoor environments, has been developed for indoor applications. This gap arises from the diverse range of use cases and their specific requirements, including cost, accuracy, deployment scale, and compatibility. Depending on the use case, certain technologies become viable. To maximize compatibility, widely standardized technologies such as \ac{Wi-Fi} and Bluetooth are often preferred.
Several studies aim to analyze the comprehensive field of indoor localization \cite{Yang2021, Zafari.2019}. However, such endeavors remain limited to a superficial level, providing taxonomies and general insights into subfields due to the broad scope and complexity of the topic.

\paragraph{Acoustic Localization}
This work focuses on acoustic localization, which is already a well-explored domain. The primary localization techniques employed include \ac{ToA} and \ac{TDoA}, with recent advancements exploring angular-based approaches. Acoustic localization enables relatively precise position estimation due to the slower propagation speed of sound in air compared to \ac{RF} signals, as demonstrated in \cite{Fischer2025}. This characteristic reduces hardware requirements and allows implementation on consumer electronics without significant modifications such as smartphones \cite{Hoflinger.2012, Fischer.2022}.

Angular-based acoustic localization represents a narrower subfield compared to TDoA and ToA methods. Systems utilizing this approach often employ microphone arrays and typically rely on chirp-like signals \cite{Gabbrielli.2023}. These signals also support multi-user identification through chirp spread spectrum techniques \cite{Gabbrielli.2023b}. Chirp signals provide robust performance against environmental disturbances and multipath fading. However, designing multi-user systems with concurrently emitting sources presents challenges, as signal collisions are difficult to resolve. Beamforming-based methods have already been employed in robotic navigation \cite{Ogiso.2019, Saqib2022} as well as the localization of walking persons\cite{Cai2021}. In the robotic context, passive, bio-inspired beacons can serve an identification function by encoding information within the structural design of a reflector \cite{Simon2020,Kroh2019}.
Orthogonal signaling schemes have also been investigated within \ac{ToA} systems to mitigate multipath interference, Doppler shifting, and related effects, while preserving multiple access capabilities \cite{Urena2018,Murano2020,Cai2020}. The conducted evaluations indicate that \ac{ZC} sequences exhibit improved robustness compared to, e.g., Kasami codes. 
A comprehensive discussion of this general topic is provided in \cite{Fischer2025}. An acoustic localization system, with multiple concurrently emitting sources, based on angular estimation, has not yet been proposed.

\paragraph{Sparse Arrays}
For several years now, undetermined estimation has been an active problem in array signal processing research \cite{Pesavento2023,Liu2023d}. Problem statements involving a minimal number of captured samples or sensor elements are usually addressed. Sparse arrays, in this context, enable the estimation of a number of sources exceeding the number of physical sensors. A comprehensive collection of algorithms, geometries, and applications is presented in \cite{Amin2024}. Specifically, sparse geometries for two-dimensional \ac{DoA} estimation are reviewed in \cite{Aboumahmoud.2021}. In the acoustic domain, several algorithms using sparse array geometries have been evaluated in \cite{Nannuru.2021}, proposing a Sparse Bayesian Learning based algorithm. In their evaluation, the authors find a combination of sparse arrays and the \ac{SBL} algorithm to perform equally with a \ac{URA} while using five times fewer sensors. Further, they show that the Co-prime arrays can resolve as many sources as there are sensors in the 2D case, while in the 1D case, they can resolve more sources than sensors. The previous work investigated several sparse geometries including Nested, Billboard, and Co-prime arrays, finding deviating performance between the geometries in an experimental setting \cite{Fischer2024b}. In particular, Nested and Billboard arrays are outperforming Co-prime arrays. However, implementation in a complete localization system has also not been shown so far.

\section{Localization Methodology}
\label{sec:loc-meth}
\begin{figure*}    
    \centering
    \includegraphics[width=1\linewidth]{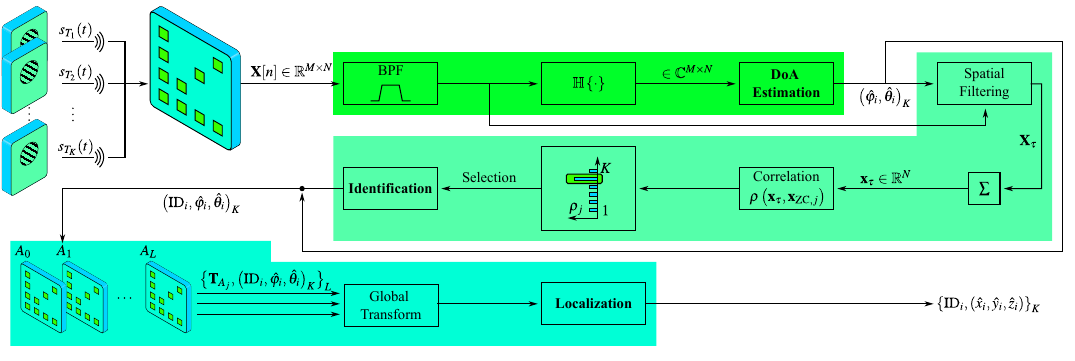}
    \caption{Block diagram of the MASSLOC signal processing chain. A multitude of $K$ sources (\(s_{T_1}(t), s_{T_2}(t), \dots, s_{T_K}(t)\)) emit orthogonal signals, which are received by microphone arrays (\(A_0, A_1, \dots, A_L\)). The received signals $\Vec{X}[n]$ undergo band-pass filtering (BPF). Direction-of-Arrival estimation computes angular information (\(\hat{\varphi}_i, \hat{\theta}_i\)), followed by identification and selection of signals through spatial filtering and correlation. The estimated IDs and DoAs are collected and transformed with their pose $\Vec{T}_{A_j}$ into a common global reference frame, in which the localization is then performed.}
    \label{fig:signal-processing-chain}
\end{figure*}

This section leads through the several parts of the localization system. Starting with a description of sparse direction-of-arrival estimation algorithms, the details of the beamforming and source identification step are covered subsequently. The last part shows how the obtained information is combined to arrive at a localization fix and in which way such systems can be calibrated.

\subsection{\markerbox{Color1} Sparse Direction-of-Arrival Estimation }
\label{ssec:sparse-loc}

Assume $K$ monochromatic sources impinge on a sparse planar array $\mathbb{S} \subset \mathbb{Z}^2$, each with a \ac{DoA} described by an azimuth angle $\varphi_i \in [-\pi, \pi)$ and an elevation angle $\theta_i \in [0, \pi/2)$ for $i = 1, \dots, K$. The received signal vector $\vec{x}_{\mathbb{S}} \in \mathbb{C}^{|\mathbb{S}|}$ is modeled as

\begin{equation}
    \vec{x}_{\mathbb{S}} = \sum_{i=1}^K A_i \vec{v}_{\mathbb{S}}(\bar{\theta}_i, \bar{\varphi}_i) + \vec{n}_{\mathbb{S}}, 
\end{equation} 
where $(\bar{\theta}_i, \bar{\varphi}_i)$ are the normalized direction cosines corresponding to the physical angles $(\theta_i, \varphi_i)$, defined as
\[
\bar{\theta}_i = \frac{d}{\lambda} \sin(\theta_i) \cos(\varphi_i), \quad \bar{\varphi}_i = \frac{d}{\lambda} \sin(\theta_i) \sin(\varphi_i),
\]
with $d$ as the grid size and $\lambda$ as the specific wavelength of the signals.
The steering vector $\vec{v}_{\mathbb{S}}(\bar{\theta}_i, \bar{\varphi}_i)$ has entries of the form

\begin{equation}
\label{eq:steering-vector}
\left[ \vec{v}_{\mathbb{S}}(\bar{\theta}_i, \bar{\varphi}_i) \right]_n = \exp\left( j 2\pi (m_{x,n} \bar{\theta}_i + m_{y,n} \bar{\varphi}_i) \right),
\end{equation}
for each sensor location $n$ in the array with coordinates $(m_{x,n}, m_{y,n}) \in \mathbb{S}$. The source amplitudes $\{A_i\}_{i=1}^K$ and the additive noise vector $\vec{n}_{\mathbb{S}}$ are modeled as zero-mean and mutually uncorrelated, satisfying
\[
\mathbb{E}[A_k^* A_\ell] = \sigma_k^2 \delta_{k,\ell}, \quad \mathbb{E}[\vec{n}_{\mathbb{S}} \vec{n}_{\mathbb{S}}^H] = \sigma^2 \vec{I}.
\]

The array covariance matrix \cite{Pal.2010} is thus 
\begin{align}
\vec{R} = \mathbb{E}[\vec{x}_{\mathbb{S}} \vec{x}_{\mathbb{S}}^H] &= \sum_{i=1}^K \sigma_i^2 \vec{v}_{\mathbb{S}}(\bar{\theta}_i, \bar{\varphi}_i) \vec{v}_{\mathbb{S}}^H(\bar{\theta}_i, \bar{\varphi}_i) + \sigma^2 \vec{I}, \\
\nonumber &\quad \vec{R} \in \mathbb{C}^{|\mathbb{S}| \times |\mathbb{S}|}.
\end{align}

The fundamental limitation in resolving multiple sources lies in the number of available eigenvalues, which corresponds to the dimension of the covariance matrix. For the \ac{MUSIC} algorithm \cite{Schmidt.1986}, this imposes a maximum number of resolvable sources satisfying $K < M$, where $M$ is the number of available sensors. This constraint motivates the use of sparse array processing. By appropriately augmenting the covariance matrix, which increases its dimension, the number of resolvable sources can be raised.

To proceed, we define the co-array of a general two-dimensional planar array geometry. Let $\mathbb{S}$ be the set of sensor positions, where each sensor is located at a 2D coordinate $\vec{p}_i \in \mathbb{Z}^2$. The co-array is then defined as the set
\[
\mathbb{D} = \left\{ \vec{m} \,\middle|\, \vec{m} = \vec{p}_1 - \vec{p}_2,\ \forall \vec{p}_1, \vec{p}_2 \in \mathbb{S} \right\}.
\]

We are particularly interested in the \textit{hole-free} subset of this co-array, denoted $\mathbb{U} \subseteq \mathbb{D}$. On an integer grid, let $\min\;\mathbb{D}$ and $\max \;\mathbb{D}$ represent the coordinate-wise minimum and maximum vectors in $\mathbb{D}$. The hole-free subset $\mathbb{U}$ is defined as the set of all grid points such that every position in the range $\left[\,\min\; \mathbb{D},\; \max\;\mathbb{D}\,\right]$ is included \cite{Amin2024}. 

\begin{figure}
    \centering
    \includegraphics[width=1\linewidth]{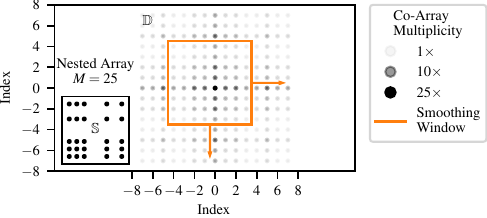}
    \caption{Co-Array of a Nested Array with $M$ sensors and the spatial smoothing window}
    \label{fig:illustration-co-array}
\end{figure}

This selection of virtual elements is crucial, since spatial smoothing can only be applied on a continuous set of differences \cite{Pal2011}. Therefore, designing sparse array geometries that maximize the size of the hole-free subset $\mathbb{U}$ is advantageous, as it enlarges the spatially smoothed (ss) covariance matrix $\vec{R}_{ss}$. In the best case, where $\mathbb{U} = \mathbb{D}$, the co-array forms a virtual \ac{URA}, which maximizes the number of resolvable sources. \autoref{fig:illustration-co-array} displays this case for a 2-level Nested Array.

\paragraph{Spatial Smoothing}
By enhancing the number of physical elements through sparse array design, the effective number of sources in the virtual array can be increased. However, this also causes the sources to appear fully coherent in the co-array domain, leading to a rank-deficient co-array covariance matrix $\vec{R}_{\text{Co}}$ \cite{Pal.2010}. Rank deficiency due to coherent sources is a phenomenon well documented in the DoA estimation literature and is especially relevant under multipath conditions, where signal copies arrive from similar directions and create strong correlations. A standard approach to mitigate this problem is to apply spatial smoothing, which aims to decorrelate the coherent sources and restore the rank of the covariance matrix.

In the one-dimensional setting, spatial smoothing is commonly implemented as a sliding window over the virtual \ac{ULA}. In the two-dimensional case, where the virtual co-array forms an \ac{URA}, the same concept is extended by averaging over overlapping rectangular subarrays. Let the virtual URA span dimensions $M_x \times M_y$ on a 2D integer grid. To estimate the normalized direction cosines $\{(\bar{\theta}_i, \bar{\varphi}_i)\}_{i=1}^K$, spatial smoothing is applied over the virtual URA co-array $\vec{x}_{\mathbb{S}_{\text{diff}}^{\text{URA}}}$ to compute the spatially smoothed covariance matrix $\vec{R}_{ss}$, which is defined as
\[
\vec{R}_{ss} = \frac{1}{L_x L_y} \sum_{p=0}^{L_x - 1} \sum_{q=0}^{L_y - 1} \vec{J}_{p,q} \vec{x}_{\mathbb{S}_{\text{diff}}^{\text{URA}}} \vec{x}_{\mathbb{S}_{\text{diff}}^{\text{URA}}}^H \vec{J}_{p,q}^H, \tag{6}
\]
where $\vec{J}_{p,q}$ is a 2D selection matrix that extracts a subarray of size $L_x \times L_y$ starting at offset $(p,q)$. This matrix is defined as
\[
\vec{J}_{p,q} = \vec{J}_p^{(x)} \otimes \vec{J}_q^{(y)}, \tag{7}
\]
with $\vec{J}_p^{(x)} \in \{0,1\}^{L_x \times M_x}$ and $\vec{J}_q^{(y)} \in \{0,1\}^{L_y \times M_y}$ being standard 1D selection matrices along the $x$ and $y$ dimensions, respectively (cf. \autoref{fig:illustration-co-array}). The operator $\otimes$ denotes the Kronecker product.

\begin{figure}
    \centering
    \includegraphics[width=1\linewidth]{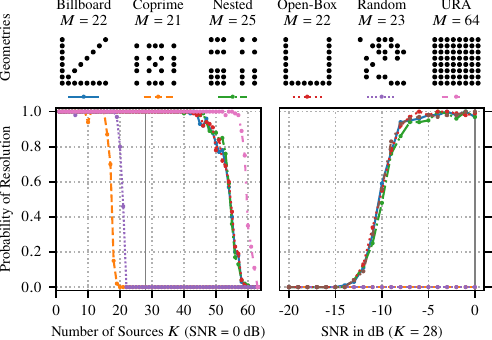}
    \caption{Resolution capabilities of various array geometries (top) over the number of sources $K$ and over the SNR (bottom).}
    \label{fig:sim-array-geometries}
\end{figure}

This 2D smoothing approach preserves the structure of the virtual URA and ensures that the resulting covariance matrix $\vec{R}_{ss}$ has sufficient rank to enable high-resolution DoA estimation using algorithms such as MUSIC. The choice of subarray size $(L_x, L_y)$ determines the amount of smoothing applied and must be selected to balance resolution and statistical reliability.
In \autoref{fig:sim-array-geometries}, the results of a simulation across several geometries are plotted. The resolution criterion, which has been applied in the computation, is a spherical error under \SI{10}{\degree}. Besides the Coprime and Random arrays, which are not able to resolve more sources than sensors, the other geometries perform on par with each other.

\subsection{\markerbox{Color6} Beamforming and Source Identification}
\label{ssec:beam-and-ident}
The signal processing chain shown in Fig. \ref{fig:signal-processing-chain} comprises two stages. In the first stage, direction estimation is performed for several sources. In the second stage, these directions are used to identify the sequences of the emitted signals.

Initially, the signal is passed through a bandpass filter and then a frequency estimation is performed to determine the center frequency of the signal matrix $\mathbf{X}\in\mathbb{R}^{M\times N}$. This data matrix is made up of the signals from the $M$ microphones and $N$ samples. The filtered signal is then converted by a Hilbert transform into a complex, analytical signal, on the basis of which the direction estimation is performed with the classical MUSIC algorithm \cite{Schmidt.1986}. This provides azimuth and elevation angle pairs $\left(\varphi_i, \theta_i\right)_K$.

In the second stage, these angle pairs are used to parameterize a spatial filter. This filter delays the signals according to the angle of incidence and the position of the microphones.
The time shift experienced by the microphone \(M_j\), which is located at the position \(\mathbf{p}_j = \left(x_j, y_j, z_j\right)\), for an incoming signal from the direction $\left(\varphi_i, \theta_i\right)$, can be expressed as
\begin{equation}
    \tau_{j} = \frac{1}{c} \mathbf{p}_j 
    \begin{pmatrix}
        \sin\theta_i \cos\varphi_i \\  
        \sin\theta_i \sin\varphi_i \\
        \cos\theta_i
    \end{pmatrix}
\end{equation}
with the speed $c$ of the transmission medium (cf. \autoref{eq:steering-vector}). 

Consider now an array of sensors, as defined above, receiving a source signal \( S(\omega) \) arriving from a direction corresponding to delays \( \tau_m \). The signal at the \( m \)-th sensor is modeled as
\[
X_m(\omega) = S(\omega)e^{-j\omega \tau_m} + N_m(\omega),
\]
where \( N_m(\omega) \) denotes the noise in frequency space.

To align the sensor outputs, each received signal is phase-compensated by applying the factor \( e^{j\omega \tau_m} \):
\[
\tilde{X}_m(\omega) = e^{j\omega \tau_m}X_m(\omega) \approx S(\omega) + e^{j\omega \tau_m}N_m(\omega).
\]

The beamformer output is then obtained by summing the phase-compensated signals:
\[
Y(\omega) = \sum_{m=1}^{M} \tilde{X}_m(\omega) = \sum_{m=1}^{M} e^{j\omega \tau_m}X_m(\omega).
\]
This coherent summation process, called \ac{DS} beamforming, reinforces \( S(\omega) \) while attenuating unaligned noise, which enables the source identification.

Next, the beamformer output \(y(t)\) is correlated with a set of \(P\) candidate test sequences \(\{s_i(t)\}_{i=1}^{P}\) (for example, complementary Zadoff–Chu sequences). For each candidate sequence $s_i(t)$, the correlation is computed as
\[
R_i = \max_{\Delta t} \left| \int y(t)\, s_i^*(t-\Delta t)\, dt \right|,
\]
where the maximization over \(\Delta t\) accounts for any time misalignment. Organizing these correlation scores for \(K\) observed signals (with \(P > K\)) forms the confidence matrix
\[
\mathbf{C} \in \mathbb{R}^{K \times P}.
\]

Finally, signal identification is accomplished by solving a linear sum assignment problem. Let \(\pi:\{1,\ldots,K\} \to \{1,\ldots,P\}\) denote the assignment function that maps each observed signal to a candidate sequence. The optimal assignment is found by solving
\[
\max_{\pi} \sum_{k=1}^{K} C_{k,\pi(k)},
\]
subject to the constraint that \(\pi\) is injective (each candidate sequence is assigned to at most one signal). This problem is efficiently solved using the Hungarian algorithm, which yields the most likely identification of the sequences corresponding to the beamformed signals. The assignment problem could also be addressed by brute force with a complexity of $O(n!)$ for square matrices. However, this becomes intractable already for smaller $n$. The derivatives of the Hungarian algorithm usually attain $O(n^3)$ \cite{Crouse2016}.

\subsection{\markerbox{Color2} Localization and Self-Calibration}
\label{ssec:loc-and-self-calib}
The last step of the localization engine is the fusion of angular measurements with each other and obtaining a calibrated pose for each array device.

\paragraph{Angular-based Localization}
\label{par:angluar-loc}
The DoA-based localization algorithm fuses direction-of-arrival measurements from multiple devices, each providing its pose (position $\mathbf{p}_i$ and orientation $\mathbf{R}_i$) and a local DoA. For each device, the local DoA is transformed into the global coordinate system using its rotation and translation, yielding a unit direction vector $\mathbf{d}_i$ in the global coordinate frame and the corresponding device position. A line is then defined for each device as 
\[
\mathbf{r}_i(t) = \mathbf{p}_i + t\,\mathbf{d}_i,
\]
where \(\mathbf{p}_i\) is the device position and \(\mathbf{d}_i\) is the global DoA vector (cf. \autoref{fig:system-model}). The source location is estimated as the point that minimizes the sum of squared distances to these lines by solving the linear system 

\begin{equation}
    \vec{A}\,\mathbf{x} = \vec{b},    
\end{equation}
with 
\[
\vec{A} = \sum_{i=1}^{L} \left( \mathbf{I} - \mathbf{d}_i \mathbf{d}_i^T \right) \quad \text{and} \quad \vec{b} = \sum_{i=1}^{L} \left( \mathbf{I} - \mathbf{d}_i \mathbf{d}_i^T \right) \mathbf{p}_i.
\]
The degree to which this equation is satisfied can be used to assess the quality of the localization estimates. The divergence metric $\mathcal{D}$ is defined here as the mean orthogonal distance between the location estimate $\hat{\mathbf{x}}$ and the closest point as projected by the DoA vector of each device, i.e.
\begin{equation}    
\label{eq:divergence-metric}
\mathcal{D}
    \;=\;
    \frac{1}{L}\,
    \sum_{i=1}^{L}
    \left\|
        \left(\mathbf{I}-\mathbf{d}_i\mathbf{d}_i^{\mathsf T}\right)
        \left(\hat{\mathbf{x}}-\mathbf{p}_i\right)
    \right\|_2
\end{equation}

\paragraph{Self-calibration}
\label{par:self-calib}
As mentioned in the introduction, the setup of an angular-based localization system requires the measurement of a total of six variables per anchor node, which can be infeasible. In the current system, this is addressed by utilizing a self-calibration algorithm.

The self-calibration algorithm jointly estimates the unknown 3D position and orientation of Device~2, along with the source-to-device distances, by exploiting the geometric consistency of direction measurements. Assume Device 1 is fixed at the origin (\(\mathbf{p}_1 = \bm{0}\)) with a known orientation (\(\mathbf{R}_1 = \mathbf{I}\)), while Device 2 has an unknown position \(\mathbf{p}_2\) and rotation \(\mathbf{R}_2\). At each time step \(t\), both devices measure unit direction vectors \(\mathbf{d}_1(t)\) and \(\mathbf{d}_2(t)\) toward a moving source whose position is denoted by \(\mathbf{s}(t)\). The source position can be expressed from each device's perspective as
\[
\mathbf{s}(t) = \mathbf{p}_1 + \mathbf{R}_1\, \mathbf{d}_1(t)\lambda_1(t) \quad \text{and} \quad \mathbf{s}(t) = \mathbf{p}_2 + \mathbf{R}_2\, \mathbf{d}_2(t)\lambda_2(t),
\]
where \( \lambda_1(t), \lambda_2(t) \in \mathbb{R}_{+} \) are unknown scalar distances from the source to Device~1 and Device~2, respectively.

Equating these two representations yields a residual error:
\[
\mathbf{r}(t) = \mathbf{d}_1(t)\lambda_1(t) - \bigl(\mathbf{p}_2 + \mathbf{R}_2\, \mathbf{d}_2(t)\lambda_2(t)\bigr),
\]
and the overall objective is to minimize the sum of squared errors over all time steps:
\[
E = \sum_{t=1}^{T} \| \mathbf{r}(t) \|^2.
\]

A known distance \(D_{12}\) between the devices (with \(\|\mathbf{p}_2\| = D_{12}\)) is imposed to resolve the scale ambiguity. Further, the direction from Device~1 to Device~2 is applied as a soft constraint with cost penalty to enhance the convergence properties. Additional constraints, such as the orthonormality of \(\mathbf{R}_2\), are enforced during the optimization process, which is typically performed via nonlinear least squares methods \cite{Shames2013,Gburrek2021}.

\section{Spectral Balancing}
\label{sec:sec-balancing}

\ac{IQ}-modulated \ac{ZC} sequences are asymmetrical around the carrier frequency, leading to distortions in the direction-of-arrival estimation. To mitigate these distortions, the waveform must be balanced by extending the previously studied ZC sequences.
Spectral balancing can be achieved through various methods. In this contribution, we investigate two approaches: the Mirror-Symmetric Zadoff-Chu (MS-ZC) sequence and the  Self-Complementary (SC) sequence. The band-overlapping method is implemented by adding the complex conjugate of the original sequence, resulting in a real-valued signal but sacrificing the constant amplitude property. In contrast, the band-separated method is obtained by appending the complex conjugate of the Fourier transform in the upper half of an unoccupied band, followed by an inverse transform.

As shown later, these sequences do not strictly preserve all properties of standard ZC sequences. However, they exhibit comparable auto- and cross-correlation characteristics, making them viable options for spectral balancing.

\begin{figure}
    \centering
    \includegraphics{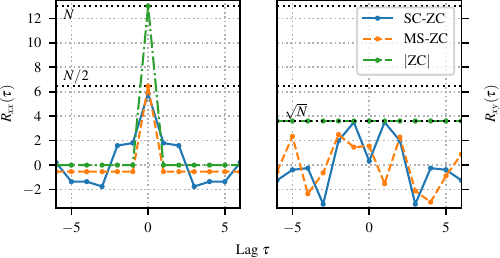}
    \caption{Exemplary cyclic auto-correlation $R_{xx}(\tau)$ (left) and cross-correlation $R_{xy}(\tau)$ (right) functions for sequences with length $N=13$ and $q_i \in \left\{1,2\right\}$.}
    \label{fig:c-zc-auto-corr}
\end{figure}

\paragraph{The Standard ZC Sequence and its Properties}

Let $N$ be a prime number and let $q$ be an integer (i.e., the \textit{root index}) satisfying $\gcd(q,N)=1$. One common definition (for odd $N$) of the Zadoff–Chu sequence is (see e.g., \cite{Beyme.2009}:
\[
x[n] = \exp\!\Biggl(-j\,\frac{\pi\,q\,n(n+1)}{N}\Biggr),\quad n=0,1,\ldots,N-1.
\]
This sequence is a \emph{CAZAC} sequence (constant amplitude zero autocorrelation) and exhibits ideal cyclic autocorrelation properties, i.e., the autocorrelation function is $N\delta[n]$. Further, its normalized cyclic cross correlation, for two sequences with the same length but different root indices, $q_i$ is exactly $1/\sqrt{N}$.

\paragraph{The Mirror-Symmetric ZC Sequence (MS-ZC)}
The first modification to arrive at an spectrally balanced sequence is to mirror the Fourier components of the standard ZC sequence in such a way that the resulting sequence is real valued.
Let $X[k]$ be the $N$-point Discrete Fourier Transform (DFT) of $x[n]$, i.e.,
\[
X[k] = \sum_{n=0}^{N-1} x[n]\,\mathrm{e}^{-j\frac{2\pi}{N} kn},\quad k=0,1,\ldots,N-1.
\]
To construct a composite sequence whose spectrum is separated into two non-overlapping bands, define a $2N$-point frequency-domain sequence, which satisfies the Hermitian symmetry
\begin{equation}    
X_\mathrm{MS}(k) =
    \begin{cases}
    0, & k = 0, \\[2mm]
    X(k), & 1 \leq k \leq N-1, \\[2mm]
    X^*(2N-1-k), & N \leq k \leq 2N-2.
    \end{cases}
\end{equation}

Taking the inverse DFT (IDFT) of $X_\mathrm{MS}[k]$, we obtain the time-domain sequence
\begin{equation}
    x_\mathrm{MS}[n] = \frac{1}{2N}\sum_{k=0}^{2N-1} X_\mathrm{MS}[k]\,\mathrm{e}^{j\frac{2\pi}{2N}kn},\quad n=0,1,\ldots,2N-1,    
\end{equation}
which can be expressed as
\begin{equation}
x_\mathrm{MS}[n] = \frac{2\sqrt{N}}{2N-1} \sum_{k=1}^{N-1} \cos\left( \frac{2\pi k n}{2N-1} - \theta(k) \right),    
\end{equation}
where the phase term \( \theta(k) \), assuming $N$ is prime, is defined as:
\[
\theta(k) = \pi \frac{q^{-1}}{N} k^2 + 1.
\]
Here, \( q^{-1} \) denotes the multiplicative inverse of $q\mod N$\cite{Beyme.2009}.

\paragraph{The Self-Complementary ZC Sequence (SC-ZC)}

An alternative approach is to form a complementary sequence directly in the time domain by overlapping the ZC sequence with its complex conjugate:
\begin{equation}
x_\mathrm{SC}[n] = \frac{1}{2}\Bigl(x[n] + x^*[n]\Bigr)
= \cos \Biggl(\frac{\pi\,q\,n(n+1)}{N}\Biggr).    
\end{equation}

This is the Self-Complementary ZC sequence since the positive and negative frequency components overlap in the spectrum, yielding a real-valued, balanced signal.

\begin{figure}
    \centering
    \includegraphics{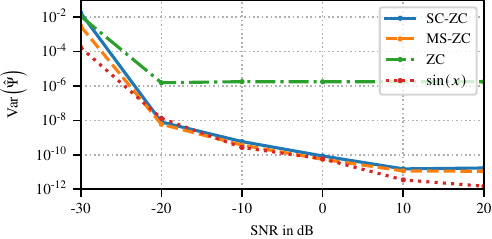}
    \caption{DoA Estimation variance of the ZC and derived sequences with a narrowband sine function as baseline.}
    \label{fig:c-zc-comparison}
\end{figure}

\paragraph{Correlation properties}
It is well known that the cyclic auto-correlation of the ZC sequence is $N\delta[n]$ (e.g. \cite{Chu1972}), where $\delta[n]$ is the dirac delta function, as shown in \autoref{fig:c-zc-auto-corr}. Evaluating the auto-correlation of the MS-ZC sequence yields a non-perfect function with a magnitude of $\frac{N(2N-2)}{2N-1}$ at $\tau=0$ and $-\frac{1}{2N-2}$ elsewhere\footnote{This result can be obtained by applying the Wiener–Khinchin theorem to the MS model in frequency space.}. The analytical structure of the SC-ZC makes the evaluation more complex. At the center, the magnitude of the auto-correlation function is $R_{xx,\mathrm{SC}} \approx N/2$. The cross-correlation function is on the order of $O(\sqrt{N})$ for all sequences (and exactly $\sqrt{N}$ for the ZC sequence). The cross-correlation function for both functions can be rephrased as two sums by the product-to-sum identity, which yields a quadratic Gauss sum \cite{Berndt1998,Gregoratti02.11.2023}.

\paragraph{Application to Direction Finding}

Based on the previously established properties of the sequences, the ZC sequence offers superior performance in terms of auto- and cross-correlation functions. However, due to its quadratic phase, it exhibits a chirp-like behavior in the frequency domain. This characteristic can be problematic for narrowband direction finding, where the phase delay is estimated for a specific carrier frequency $f_c$. If the averaging window precisely matches the length of one ZC symbol, assuming continuous transmission, this issue will not arise. However, in scenarios requiring higher time resolution, such as high-speed tracking, a shorter averaging window may be used, capturing only a subsection of the ZC symbol. This results in a skewed mean frequency within the window, introducing a systematic bias in the estimated direction vector, as illustrated in \autoref{fig:c-zc-comparison}.
The figure presents the variance of the estimated direction angles, showing that for standard ZC sequences, an increased SNR does not necessarily lead to reduced estimation variance. In contrast, spectrally-balanced versions can mitigate this effect. In general, the higher the elevation angle, the greater the variance introduced by frequency mismatch.

\paragraph{Source Identification Performance}
\begin{figure}
    \centering
    \includegraphics{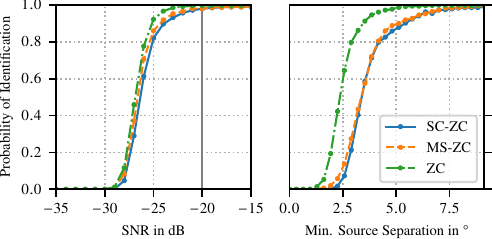}
    \caption{Identification performance with standard system parameters for $K=15$ sources over a SNR range (left) and over a minimum source separation at \SI{-20}{\deci\bel} SNR (right).}
    \label{fig:c-zc-identification}
\end{figure}
The identification performance of different sequences is influenced by their auto- and cross-correlation properties, leading to varying results under normal system operation. To assess this, a simulation was conducted, varying both the SNR and the minimal source separation. The parameters of the simulation were chosen to be the same as the system parameters stated later in \autoref{sec:exp}.
As shown in \autoref{fig:c-zc-identification}, the performance differences between the sequences remain small across different SNR levels. However, a closer analysis confirms that the ZC sequence performs best, as expected. The simulation includes a total of $K=15$ sources, with the minimal source separation angle varied at an SNR of \SI{-20}{\deci\bel}. When beamforming is applied, the ZC sequence can identify sources that are closer together compared to the other sequences. In contrast, the MS and SC sequences exhibit similar identification performance, without significant differences between them.
\section{System Overview and Hardware Concept}
\label{sec:system-overview}
The system concept, illustrated in Fig. \ref{fig:system-model}, consists of $K$ tags, each emitting a unique \ac{ZC} sequence.

The proposed array device is based on a 64-microphone \ac{URA} system. The utilized digital microphones are of the type TDK ICS-52000 (TDK InvenSense Inc., USA). Further, the array module integrates a Cyclone V \ac{FPGA} (Altera Corp., USA), which includes a \ac{HPS}. The collected data is transmitted via Ethernet to a central localization server, where \ac{DoA} estimation as well as identification and localization, are performed. In reference \cite{Fischer2024b}, a comprehensive description of the complete array device alongside performance measures is given.

Each tag is equipped with a SP1303L dynamic speaker (Soberton Inc, USA) and an embedded system based on an EFM (Silicon Laboratories, Inc, USA) microcontroller. The tags are preloaded with predefined signal waveforms stored in flash memory. A \SI{868}{\mega\hertz} link is available to control the tags, enabling functions such as starting and stopping sound broadcasts.
\section{Experimental Evaluation}
\label{sec:exp}
\begin{figure}
	\begin{centering}
		\includegraphics[width=1\linewidth]{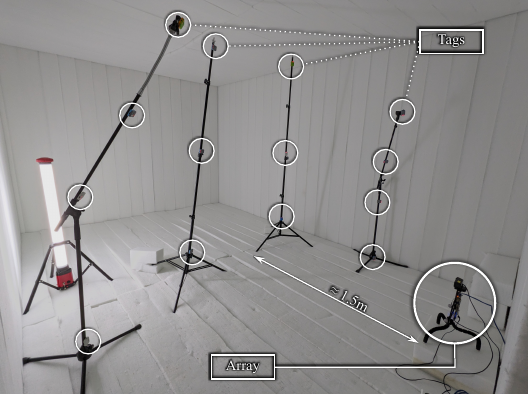}
		\caption{Anechoic Chamber setup with a multitude of 14 acoustic tags.}
		\label{fig:anechoic-chamber}
	\end{centering}
\end{figure} 

The system is evaluated in two different settings. First, we demonstrate the system's capabilities in an anechoic chamber setting. Then, the complete localization system is evaluated in a challenging, reverberant indoor environment. The following experiments are performed with a ZC symbol time $T_{ZC}=\SI{0.5}{\second}$, a bandwidth of $B=\SI{1.5}{\kilo\hertz}$ at a carrier frequency of $f_c=\SI{18}{\kilo\hertz}$. For direction estimation, a chunksize $N_c=\num{4096}$ at a sample rate of $f_s=\SI{48828}{\hertz}$ is assumed, chunks are shifted at a rate of \SI{100}{\hertz}. If not stated otherwise, the standard \ac{URA} will be used with the MUSIC algorithm to obtain direction estimates.

\subsection{Multi-Source Evaluation in an Anechoic Chamber}
\begin{figure}
	\begin{centering}
		\includegraphics[width=1\linewidth]{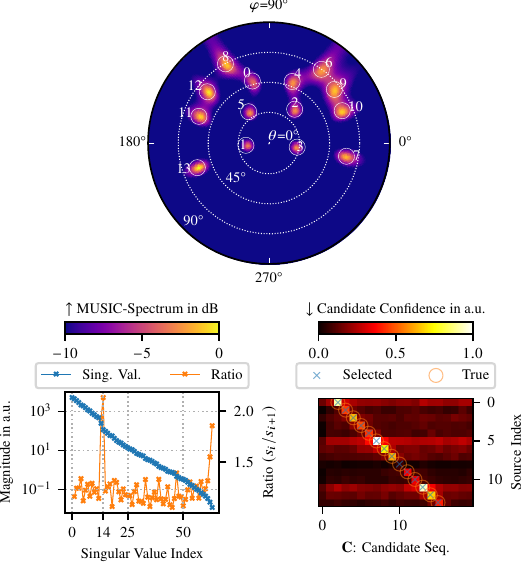}
		\caption{Multi-source experiment conducted in an anechoic chamber. MUSIC pseudospectrum (top), illustrating the estimated source directions, the singular values of the data matrix and their successive ratios (left) and the confidence matrix~$\mathbf{C}$ (right), comparing the identification confidence of candidate sequences.}
		\label{fig:multisource-composite-anechoic}
	\end{centering}
\end{figure}

The first experiment demonstrates the capability of the system to estimate the direction-of-arrival and identify a large number of simultaneously active sources. The measurements were conducted in an anechoic chamber (Wendt-Noise Control GmbH, Germany). The chamber exhibits a reverberation time ($T_{60}$) of approximately \SI{15}{\milli\second}, which is generally considered very low in the context of room acoustics \cite{Kuttruff2017}. A total of 14 tags were placed on four separate tripods, as illustrated in \autoref{fig:anechoic-chamber}, with each tag emitting a unique SC-ZC sequence. The results of the experiment are shown in \autoref{fig:multisource-composite-anechoic}.

The number of active sources was correctly estimated using the ratio of singular values, and the \acp{DoA} were obtained via the MUSIC pseudospectrum. Additionally, the candidate confidence matrix, which is obtained through beamforming and evaluated against a pool of 20 candidate sequences, is shown. Using the assignment algorithm described in \autoref{ssec:beam-and-ident}, all 14 sources are identified correctly.

\subsection{Real-World Evaluation in a Reverberant Lobby Environment}

\begin{figure}
	\begin{centering}
		\includegraphics[width=1\linewidth]{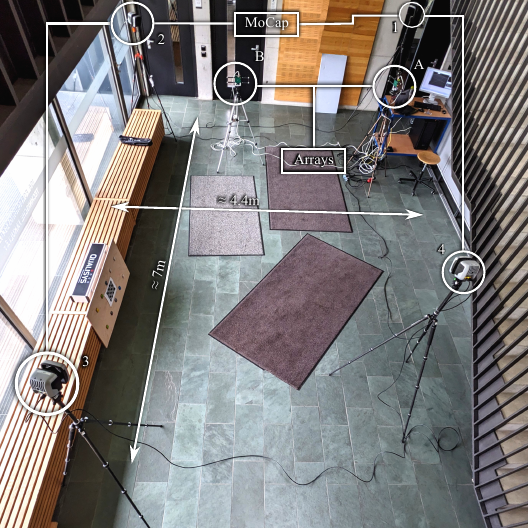}
		\caption{Evaluation environment in lobby with experimental setup.}
		\label{fig:foyer-setting}
	\end{centering}
\end{figure}

In a subsequent experiment, we assessed the localization and identification performance of the system in an indoor test environment. The setup, illustrated in \autoref{fig:foyer-setting}, covers an area of approximately \SI{31}{\meter\squared}. Two microphone arrays were installed in the university lobby, accompanied by four \ac{MoCap} cameras (Arqus A12, Qualisys AB, Sweden) operating at a capture rate of \SI{100}{\hertz}. Both the arrays and the tags were equipped with spherical tracking markers to enable accurate ground truth position estimation. The lobby is primarily constructed from natural stone slabs and concrete, with a large glass wall on one side. Additionally, some wooden panels and carpets are distributed across the floor. This combination of materials results in a measured reverberation time ($T_{60}$) of approximately \SI{1.6}{\second}, which is characteristic of acoustically live environments such as concert halls or opera houses. In contrast, office spaces are generally designed to exhibit much shorter $T_{60}$ values, typically below one second~\cite{Kuttruff2017}. For example, the acoustic laboratory in front of the anechoic chamber, which is comparable to an office space, exhibits a $T_{60}$ value of around \SI{600}{\milli\second}. Overall, higher $T_{60}$ values tend to negatively impact localization performance due to increased multipath propagation and a higher baseline noise floor.

\paragraph{Localization Accuracy for a Single Moving Tag}

\begin{figure}
	\begin{centering}
        \includegraphics[width=1\linewidth]{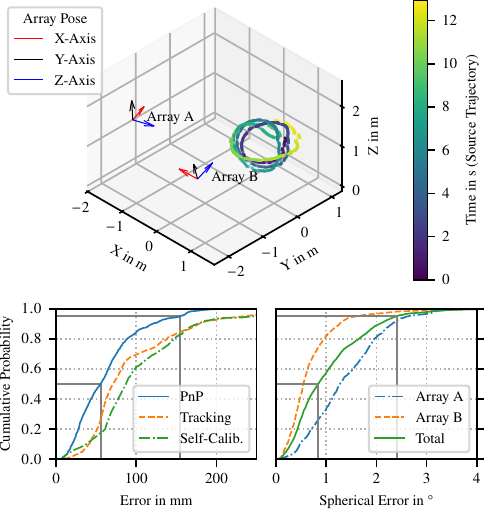}
		\caption{Single Source Performance Evaluation. The tag trajectory (top) is shown with the cumulative error probability of position (left) and spherical errors (right). }
		\label{fig:eva-single-source}
	\end{centering}
\end{figure}

In this experiment, a tag emitting an SC-ZC sequence with ID 10 was moved throughout the lobby space with a median velocity of $\approx\SI{1}{\meter\per\second}$ (with a 95\% range of \SIrange{0.1}{1.9}{\meter\per\second}). Accurate estimation of the arrays' poses is crucial for achieving high localization performance. Three methods for determining the array poses were evaluated. The first method (\textit{Tracking}) uses the positions inferred from the MoCap system by tracking spherical markers attached to the arrays. The second method relies solely on the ground truth positions of the tag, as measured by the MoCap system, and determines the array positions by solving an optimization problem based on the measured DoAs and the known tag positions. This approach corresponds to the classical Perspective-n-Point (\textit{PnP}) problem \cite{Hahmann2022,Bishop2009}. Notably, this method could be applied in practical setups without a MoCap system by placing the tag at several known positions and recording the directional data from multiple microphone arrays. \textit{Self-calibration} refers to the third method, as discussed in \autoref{ssec:loc-and-self-calib}, which is the most convenient in the setup and installation process. The cumulative error distribution plots, as well as the quartile metrics, are presented in \autoref{fig:eva-single-source} and \autoref{tab:spherical-position-metrics}. The three methods exhibit different \ac{CDF} curves, with the PnP method being the most accurate. Self-calibration falls in a comparable range as the Tracking method, yet there is a premium in terms of localization accuracy to be paid for the ease of use. The velocity exhibits only a weakly negative Pearson correlation ($\rho_{XY}=-0.3$) with the positional error, which indicates some robustness in the positional estimation of moving sources.

\begin{table}
    \centering
    \renewcommand{\arraystretch}{1.2}%
    \begin{tabularx}{\linewidth}{@{}lXXX@{}}\toprule
         Error Metric (Quartile) & 50\% & 95\% & 100\% \\
         \midrule
         Position (PnP) in mm & \textbf{55.68} & \textbf{154.34} & \textbf{228.67} \\         
         Position (Tracking) in mm & 71.01 & 236.62 & 389.63 \\ 
         Position (Self-Calibration) in mm & 86.00 & 244.14 & 283.81 \\
         \midrule
         Spherical (Array A) in \textdegree & 1.27 & 2.65 & 4.96 \\
         Spherical (Array B) in \textdegree & 0.54 & 1.64 & 4.51 \\
         Spherical (Total) in \textdegree   & \textbf{0.84} & \textbf{2.41} & \textbf{4.96} \\
         \bottomrule
    \end{tabularx}
    \caption{Comparison of error metrics for the single source performance. The positional error (left) and  the spherical error (right) with the PnP pose is shown.}
    \label{tab:spherical-position-metrics}
\end{table}

\begin{figure}
    \centering
    \includegraphics{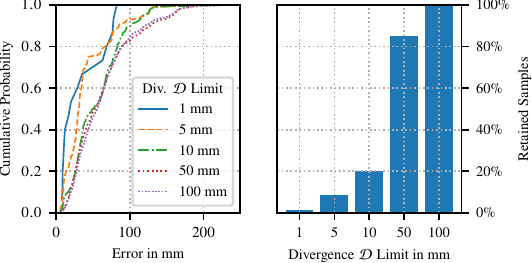}
    \caption{Accuracy–Latency Trade‑off: Effect of the divergence limit~$\mathcal{D}$ on the cumulative distribution of localization error (left) and the percentage of retained samples (right).}
    \label{fig:div-limit}
\end{figure}

The divergence limit, as defined by \autoref{eq:divergence-metric}, is a design parameter that balances localization accuracy and latency. Increasing the exclusion limit results in fewer valid measurement samples and greater spacing between them; however, the positional error decreases. \autoref{fig:div-limit} illustrates this relationship for several selected divergence limits. For example, choosing a limit of \SI{1}{\milli\meter} (instead of the default value of \SI{100}{\milli\meter}) reduces the 95\,\%‑percentile error to \SI{82}{\milli\meter}, but only about \SI{1.5}{\percent} of the samples remain valid, which is approximately one sample per second.


\paragraph{Localization Performance with Multiple Sources}

\begin{figure}
    \centering
    \includegraphics{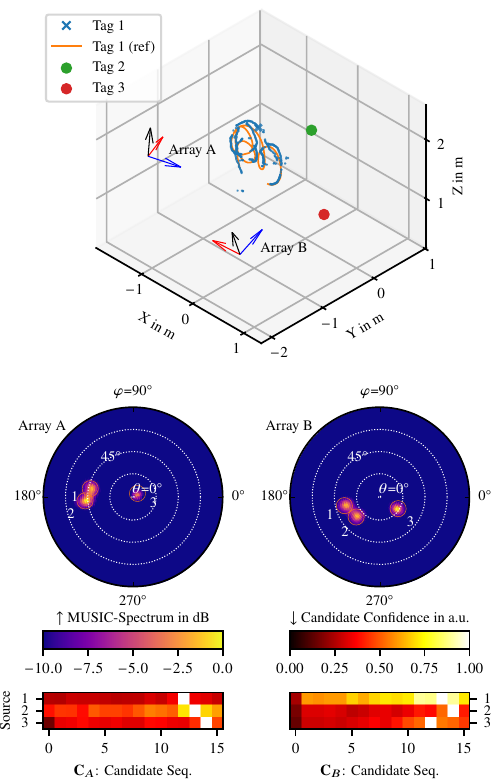}
    \caption{Multi-Source Experiment: Setup with three tags (where Tag 1 is dynamically moving, top). The MUSIC-Pseudospectrum with estimation marked (center) and the confidence matrices $\mathbf{C}_{A,B}$ for source identification (bottom).}
    \label{fig:multi-source-setup}
\end{figure}

\begin{figure}
    \centering
    \includegraphics{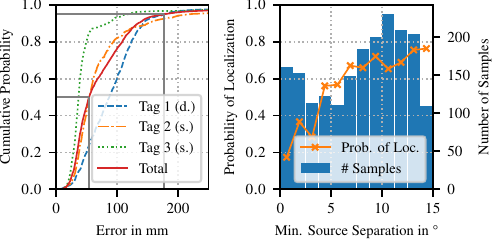}
    \caption{Multi-Source Experiment: Overall error performance (left, s.=static, d.=dynamic) and Probability of Localization (right) by the minimal spherical separation of the sources with the total number of samples shown.}
    \label{fig:multi-source-perf}
\end{figure}

To demonstrate the system's capability for concurrent localization of multiple sources, we evaluated a scenario with three simultaneously active tags in this experiment. One tag (Tag~1) was moved throughout the localization space, while the other two (Tags~2 and~3) remained at static positions. All tags continuously emitted SC-ZC sequences with IDs in the range $q_i\in\{12, 13, 14\}$. The localization process and the overall experiment setup are illustrated in \autoref{fig:multi-source-setup}.
The trajectory of Tag~1 is partially shown with its corresponding MoCap reference trajectory for validation. The MUSIC pseudospectrum visualizes the three detected sources for both arrays A and B. Following direction estimation, a beamforming process generates the source identification confidence matrices $\mathbf{C}_A$ and $\mathbf{C}_B$, which are displayed at the bottom of \autoref{fig:multi-source-setup}.
Overall, the system maintains robust performance under multi-source conditions, with only slight degradation in some accuracy metrics: the median error (50\%) is \SI{54}{\milli\meter}, the 95th percentile is \SI{178}{\milli\meter}, and the maximum observed error (100\%) is \SI{3456}{\milli\meter}.
This experiment also addresses the system’s limitation regarding minimal source separation, as shown in the right plot of \autoref{fig:multi-source-perf}. The localization quality will deteriorate significantly when sources are closely positioned, particularly at angular separations below $5^\circ$, corresponding to small spherical distances between the DoA vectors.

\paragraph{Performance of Sparse Geometries}

\begin{figure}
    \centering
    \includegraphics{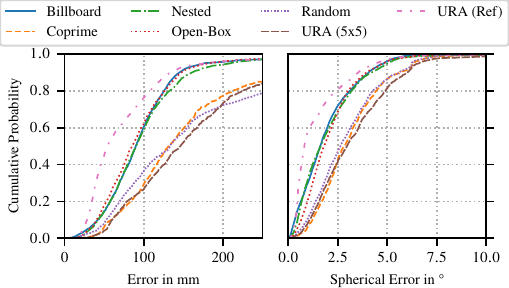}
    \caption{Multi-Source Experiment: Localization performance with Spatial-Smoothing (SS) MUSIC applied to sparse array geometries.}
    \label{fig:multi-source-sparse}
\end{figure}

The previously introduced array geometries were evaluated on the recorded data by masking the full \ac{URA} and applying spatial-smoothing MUSIC for direction estimation (cf. \autoref{ssec:sparse-loc}). As shown in \autoref{fig:multi-source-sparse}, the \ac{CDF} of localization errors are compared across all geometries, using the full URA with standard MUSIC as the reference. Detailed numerical results for each configuration, with position and angular errors at key percentiles, are provided in \autoref{tab:geometry-performance}. Among the sparse geometries, Billboard, Nested, and Open-Box arrays performed similarly well, each achieving approximately two-thirds of the valid samples obtained by the full URA. In contrast, the Coprime, Random and 5x5 reduced Rectangular ($M=\num{25}$) arrays showed the weakest performance, both in terms of localization accuracy and the number of valid measurements retained.

\begin{table}
    \centering
    \renewcommand{\arraystretch}{1.2}%
    \begin{tabularx}{\linewidth}{@{}lXXXr@{}}\toprule
     Geometry & 50\% & 95\% & 100\% & Valid Samples \\
     \midrule
     Billboard   & 90\,mm   & \textbf{179\,mm}  & 2285\,mm &  66\,\%\\
                 & \textbf{1.48°}    & 5.31°    & 61.53°   & \mybar{66}\\

     Coprime     & 132\,mm  & 323\,mm  & \textbf{1438\,mm} & 27\,\%\\
                 & 2.81°    & 7.13°    & 59.31°   &  \mybar{27}\\

     Nested      & 90\,mm   & 210\,mm  & 1832\,mm &  70\,\%\\
                 & 2.23°    & 5.58°    & 61.93°   & \mybar{70}\\

     Open-Box    & \textbf{85\,mm}   & 184\,mm  & 5318\,mm & 65\,\% \\
                 & 1.82°    & \textbf{5.19°}    & 85.20°   & \mybar{65}\\

     Random      & 133\,mm  & 409\,mm  & 1467\,mm & 28\,\% \\
                 & 3.45°    & 7.09°    & \textbf{40.86°}   & \mybar{28}\\
                 
     URA (5x5)   & 147\,mm  & 383\,mm  & 1973\,mm & 31\,\% \\
                 & 2.97°    & 6.62°    & 52.04   & \mybar{31}\\

    \midrule
     URA (Ref)   & {54\,mm} & {178\,mm} & {3456\,mm} &  100\,\% \\
                 & {0.80°}  & {4.43°}   & {56.79°}   & \mybar{100}\\
     \bottomrule
    \end{tabularx}   

    \caption{Comparison of position and spherical error metrics across different array geometries. Each geometry is listed with its median (50\%), 95th percentile, and worst-case (100\%) errors. Valid sample rates are shown as percentages and horizontal bars.}
    \label{tab:geometry-performance}
\end{table}

\section{Conclusions}
\label{sec:conclusions}
In this contribution, we proposed an acoustic, angular-based indoor localization system that utilizes sparse two-dimensional array geometries combined with complementary Zadoff–Chu sequences for accurate source localization and identification. The system and its overall concept have been evaluated both in a controlled laboratory setting and in a demanding, reverberant indoor environment. To this end, several self-calibrating algorithms have been investigated, which substantially reduce the installation effort of such a system in real-world scenarios.

The system has demonstrated the capability to fully detect, localize (in terms of direction-of-arrival estimation), and identify up to \num{14} concurrently emitting tags in a laboratory environment. Using a simple Perspective-n-Point calibration procedure, the system achieves a median 3D localization error of \SI{55.7}{\milli\meter} (95\,\%: \SI{154}{\milli\meter}) and a median spherical error of \SI{0.84}{\degree} (95\,\%: \SI{2.41}{\degree}) in a lobby with a reverberation time of \SI{1.6}{\second} ($T_{60}$). In the same environment, the accuracy with three concurrently emitting tags was evaluated to yield a median position error of \SI{54}{\milli\meter} and a 95\,\% error of \SI{178}{\milli\meter}, with performance degrading significantly only when the angular separation drops below \SI{5}{\degree}. By reducing the divergence limit from \SI{100}{\milli\meter} to \SI{1}{\milli\meter}, the 95\,\% error is halved (from \SI{154}{\milli\meter} to \SI{82}{\milli\meter}), albeit at the cost of retaining only approximately \SI{1.5}{\percent} of the samples. This trade-off may be beneficial in scenarios where highly accurate but low-dynamic localization fixes are required. Sparse geometries have also been evaluated using the captured data, revealing that Billboard, Nested, and Open-Box configurations retain approximately \SIrange{66}{70}{\percent} of valid samples with median localization errors from \SIrange{85}{90}{\milli\meter}, while using no more than \SI{40}{\percent} of the sensors. In contrast, Coprime and Random geometries, as well as the reduced URA, show clearly inferior performance. 

These results indicate that the proposed concept is scalable to a larger number of tags (greater than \num{10}) while maintaining sub-decimeter accuracy. Furthermore, through the use of self-calibrating algorithms, the system can be easily deployed in practical, real-world use cases. However, the direction-of-arrival estimation is not yet implemented in a real-time fashion, suggesting a potential future direction involving the use of the on-board FPGA's parallel processing capabilities to reduce network traffic. Source tracking has not yet been addressed; since the direction-of-arrival measurements lie on a sphere and follow a von Mises–Fisher distribution, traditional linear Kalman filter techniques may not provide sufficient accuracy to incorporate directional data effectively \cite{Mardia1999}. Future work will focus on this aspect, potentially also leveraging data augmentation to improve robustness in situations where data is only partially available.

\section*{Acknowledgments}
The authors thank Johannes Wendeberg and Frederik Riedel of Telocate GmbH for their support with acoustic equipment, and the IMBIT / BLBT at the University of Freiburg for providing access to the motion capture system.

\bibliographystyle{IEEEtran}
\bibliography{references}

\end{document}